\documentclass {article}

\usepackage[latin1]{inputenc}
\usepackage[T1]{fontenc}
\usepackage{amsmath}
\usepackage{amsfonts}
\usepackage{amssymb}
\usepackage{latexsym}
\usepackage[english]{babel}
\usepackage[dvips]{graphicx,color}
\usepackage{latexsym}
\usepackage{graphicx}
\usepackage{graphics}
\usepackage{pstricks}

\begin{document}

\newcommand{\be}{\begin{equation}}
\newcommand{\ee}{\end{equation}}
\newcommand{\epe}{\end{equation}}
\newcommand{\bea}{\begin{eqnarray}}
\newcommand{\eea}{\end{eqnarray}}
\newcommand{\ba}{\begin{eqnarray*}}
\newcommand{\ea}{\end{eqnarray*}}
\newcommand{\epa}{\end{eqnarray*}}
\newcommand{\ar}{\rightarrow}

\def\r{\rho}
\def\D{\Delta}
\def\R{I\!\!R}
\def\l{\lambda}
\def\D{\Delta}
\def\d{\delta}
\def\T{\tilde{T}}
\def\k{\kappa}
\def\t{\tau}
\def\f{\phi}
\def\p{\psi}
\def\z{\zeta}
\def\ep{\epsilon}
\def\hx{\widehat{\xi}}
\def\a{\alpha}
\def\b{\beta}
\def\O{\Omega}
\def\M{\cal M}
\def\g{\hat g}
\newcommand{\dslash}{\partial\!\!\!/}
\newcommand{\aslash}{a\!\!\!/}
\newcommand{\eslash}{e\!\!\!/}
\newcommand{\bslash}{b\!\!\!/}
\newcommand{\vslash}{v\!\!\!/}
\newcommand{\rslash}{r\!\!\!/}
\newcommand{\cslash}{c\!\!\!/}
\newcommand{\fslash}{f\!\!\!/}
\newcommand{\Dslash}{D\!\!\!\!/}
\newcommand{\Aslash}{{\cal A}\!\!\!\!/}

%\newcommand{\u}{\underline}
%\begin{center}
\hspace{12cm} CERN-PH-TH/2011-115
\vspace{3mm}

\begin{center}

\vspace{6mm}

{\large Spacetime Geometry as Statistic Ensemble of Strings}

\vspace{.3in}
Marcelo Botta Cantcheff $^{\dag\,\ddag}$\footnote{e-mail:
bottac@cern.ch, botta@fisica.unlp.edu.ar}

\vspace{.2 in}

$^{\dag}${\it CERN, Theory Division, 1211 Geneva 23, Switzerland}\\
$^{\ddag}${\it IFLP-CONICET and Departamento de F\'{\i}sica\\ Facultad de Ciencias Exactas, Universidad Nacional de La Plata\\
CC 67, 1900,  La Plata, Argentina}\\
\vspace{.4in}
\end{center}

\begin{abstract}
\noindent

Jacobson theorem (Ref. \cite{jacobson}) shows that Einstein gravity may be understood as
 a thermodynamical equation of state; a microscopic realization of this result is however lacking.
  In this paper, we propose that this may be achieved by assuming the spacetime geometry as
   a macroscopic system, whose thermodynamical behavior is described by a statistical ensemble,
    whose microscopic components are low-dimensional geometries. We show that this picture
     is consistent with string theory by proposing a particular model for the microscopic geometry,
      where the spacetime metric plays the role of an ordinary thermodynamical potential in a special ensemble.
 In this scenario, Einstein equation is indeed recovered as an equation of state,
 and the black hole thermodynamics is reproduced in a thermodynamic limit (large length scales).
The model presented here is background-independent and, in particular, it provides an alternative formulation of string theory.

\end{abstract}

\section{Introduction.}

Strong facts such as the Unruh effect and black hole physics suggest the close relation between geometry and thermodynamics.
 Ted Jacobson realized that the Einstein equation may be recovered as an equation of state from the first law of thermodynamics,
% the fundamental thermodynamic relation,
 %$\delta Q = T \delta s$ between the energy that flows across a causal horizon $Q$, temperature $T$ and entropy $s$,
and the Beckenstein-Hawking formula for the entropy of horizons \cite{jacobson}.
An open problem of theories that should describe gravity at a fundamental level (such as string theory) is precisely to explain how the Jacobson result is recovered from a microscopic description. In this sense, Einstein gravity should be recovered from the statistics of fundamental systems in a proper thermodynamical limit.

Many authors have explored the possibility of describing gravity as a macroscopic theory in many senses \cite{sak,ens,ens2}, and more recently, the ideas of holography and emergence of the spacetime have open new perspectives in this direction \cite{pad,seiberg,verlinde}.
According to this paradigm, it is often claimed that gravity should not be quantized as the other field theories that describe the matter
 and its interactions, since it should not be considered a fundamental theory but an emergent one from microscopic exactly quantizable degrees
  of freedom (``atoms of spacetime'' \cite{ens}\cite{atoms}).
%(``atoms of spacetime'', see for instance \cite{atoms}).
 However, so far, this claim has never been shown or convincingly justified from the basic principles of some concrete theory o model.
%Our first goal here is to propose a model with this property.

In the present work, we explore the possibility of considering the spacetime as a macroscopic body which may be studied according to the methods of statistical mechanics and thermodynamics. So, metric and all other tensors that characterize the geometry are thought of as thermodynamical variables.
A first step to construct a concrete model with these properties was given in Ref. \cite{ens}, where we claimed that the spacetime geometry emerges from microscopic degrees of freedom, and a particular and simple way of realizing this hypothesis was also proposed: to assume that the spacetime is composed by a large number of low-dimensional geometries where the microscopic (fundamental) degrees of freedom live on, and are described by exactly solvable theories. So in agreement with the spirit of the Jacobson paradigm, the quantities (tensors) that describe the spacetime geometry shall be recovered as statistic averages on ensembles of such systems
\footnote{This proposal refers to the nature of classical (microscopical) degrees of freedom of the spacetime as the object to be quantized, but does not constitute a quantum gravity itself, so the macroscopic spacetime $(M,g)$ should be recovered from this microscopical structure in the proper thermodynamical/macroscopical approximation rather than by taking a semiclassical limit.}.
Notice that the information paradox has an immediate resolution in this context since the quantum theory of the microscopical degrees of freedom is unitary, while the spacetime physics is described by mixed states initially, and among its time evolution.

\vspace{0.7cm}

On the other hand, there is a general agreement that string theory plays a central role in describing spacetime geometry, gravity, and also matter and interactions at the more fundamental level. A serious theoretical problem however is that the theory is defined perturbatively, and formulated in a background-dependent way. Our purpose in this paper is to propose a consistent framework where the point of view discussed above and the main features of string theory may be conciliated in a formulation where the role played by the background geometry is conceptual and quantitatively different. In particular,
we are going to assume that the microscopic constituents of the geometry are two-dimensional Lorentzian manifolds, embedded into an ambient (spacetime) manifold $M$ whose local geometry plays no role in the microscopic theory, and it just appear as averages over ensembles. In fact, the background metric will be treated as a thermodynamical variable, introduced as a standard Lagrange multiplier in the proper statistic ensemble, in the same way a chemical potential does it.
The result is that this reproduces string theory, but also has novel implications on the very meaning of the theory: it opens up new perspectives for the
 description of the background physics and the interpretation of the string states itself.

This work is organized as follows: in Section 2, we formulate the fundamental theory on microscopic geometries and discuss its relation with macroscopic geometry through statistics; in Section 3, we show how the model works in the simplest canonical ensemble where only no local aspects of the spacetime may be described, and the generalization to other ensembles is discussed. In section 4, we define the geometric ensemble, where the spacetime metric appears as a thermodynamical potential, and show that the Polyakov path integral of string theory is recovered. The Einstein equation is shown to be an equation of state in this ensemble and the black hole entropy is evaluated. Finally in Section 5, we show how to localize the constraints and generalize the construction to a non-uniform system, in order to recover the string theory as a non-linear sigma model. Concluding remarks are collected
in Section 6.

\section{The microscopic theory and spacetime as statistical concept}

The general properties and ideas on the spacetime geometry as a complex system of simplest microscopic ones discussed above were collected in a
 hypothesis introduced some years ago \cite{ens}. Here we discuss a more specific realization of this proposal and suppose that the fundamental (microscopical) theory is given by a collection of oriented two-dimensional connected differentiable manifolds $W_n$, equipped with a Lorentzian local metric $q_{ab}$, and embedding fields $X_{(n)} : W_n \to M$ into a $d$-dimensional differentiable manifold $M$. All the fundamental degrees of freedom are described by an exact unitary quantum field theory on $W_n$.
The most general set of microscopic fields may be expressed by $\z=(X^\mu(\sigma^c), q_{ab}(\sigma^c),\phi(\sigma^c))$ on $W\equiv \cup^N_n\,W_n $, whose dynamics is defined by the  action:
\be\label{action}
{\cal S}[\z]= \lambda \, {\cal S}_G \,+\, \int_{W} d^2 \sigma\,\sqrt{-q}\,\, {\cal L}_m [ q,\phi]
\ee where $\sigma^a, a=0,1$ denotes the local coordinates of $W_n$.
This is invariant under arbitrary smooth coordinate changes of $W$ ($Diff\,(W)$). We also demand conformal invariance to be a fundamental symmetry of this theory. The component $\phi(\sigma^c)$ denotes all the intrinsic fields on $W_n$, which may describe internal gauge symmetries.

 The first term is a purely geometric theory, given by the Einstein-Hilbert action
 \be\label{geometric-action} {\cal S}_G[q, W] =  \; \frac{1}{4\pi}\int_{W} d^2 \sigma\,\, \sqrt{-q} R[q]  +  \frac{1}{2\pi}\int_{\partial
W} ds\, k  ,\ee
where the second term is the usual boundary term and $k$ is the geodesic curvature of $\partial W$. There are other interesting possibilities for two-dimensional gravities which will be not analyzed in this paper: extra terms contributions (e.g. higher order terms and cosmological term), a Liouville theory, etcetera \footnote{A large number of three-dimensional manifolds as the microscopic picture of the spacetime, was discussed as example in Ref. \cite{ens}.}.

This theory describes the intrinsic geometry of $W$ completely, while the embedding target $M$, which we identify with the spacetime manifold, is assumed to be differentiable with no metric or any other \emph{a priori} geometric structure than its topology, which for simplicity we assume to be  $M \sim \Sigma \times {\mathbb R}$ where $\Sigma$ is compact (this plays the role of a "box" for the system). In what follows we shortly refer to these microscopic components of the geometry as \emph{strings}, and to $W$ as \emph{string foam} or microscopic geometry. Let us emphasize however that $W$ is not a "quantum" foam but a microscopic picture of the classical spacetime.

We require that the theory be background-independent, i.e. the quantities that characterize the local geometry of the spacetime shall emerge from the microscopic theory through a thermodynamical/macroscopic description.
In fact, the fundamental object that characterizes all the macroscopical physics, the spacetime inclusive, is the statistic operator $\r$ on the Hilbert space
 ${\cal H}_{(\z,W)}$ of the quantized microscopic theory.
%\textbf{A configuration basis of this space may be denoted by $| \z \rangle$, so then  $\r =\Sigma_\z \,\,\r(\z) | \z \rangle \langle \z | $.}
 Notice that the information paradox is automatically absent in this picture since the microscopic theory is a well defined unitary quantum field theory, and the spacetime geometry is described by mixed states encoded in $\r$, that evolves according to the rules of the statistical mechanics.
 The thermodynamical entropy is canonically defined as $s \equiv -k_B \; Tr \left(\r \;ln \,\r \right) $ (we set $k_B=1$), which must be a maximum in the equilibrium states compatible with certain macroscopic constraints. So indeed, the spacetime physics shall be derived from a maximum thermodynamical entropy principle.

Obviously, the embedding variables $X$ may be changed by means of general coordinate changes of the manifold $M$, so $Diff (M)$ (diffeomorphisms of $M$) must be a symmetry of the theory. Therefore, any diffeomorphism-invariant contribution of the field $X$ to the action (\ref{action}) requires contractions with a $M$-metric, which would violate the background independence. By virtue of this, the fundamental theory must be independent of the field $X$. This means that the theory is invariant under movements of the string foam in the spacetime manifold, and it does not describes the extrinsic geometry of $W$.

Let us conclude this section by emphasizing that, according to this viewpoint, \emph{quantum gravity} is nothing but the quantization of this two- dimensional theory. The problem of how to sum over distinct topologies in quantum gravity is radically reassessed in this approach as summing on two-dimensional geometries, which may be organized in terms of simple discrete invariant, the genus. In what follows we are devoted to show how the relevant information of the geometry of the spacetime, and spacetime fields, may be extracted from proper statistic ensembles of such objects.

%A conceptually radical fact is that the problem of how to sum over distinct topologies in quantum gravity is handled/reminded

\section{Spacetime geometry from thermodynamical ensembles}

The maximum entropy principle supposes that the time scales used by the observers to measure times are very large compared with the thermalization time scales of the microscopical system, such that we observe the space time in a thermal equilibrium state. The same assumption may be done on the length scales to get an homogeneous system. A precise notion of what these scales mean shall can be given after appearing the spacetime metric in the model.

Let us define the vector field ${\bf U} = U^\mu\frac{\partial}{\partial X^\mu}\,\, \mu=0,1,...,d-1$ in the tangent space $T_XM$ as ${\bf U}=\frac{\partial}{\partial X^0}$, which may be interpreted as the tangent vector to geodesics described by macroscopic static observers in each space point. For clearness, it is convenient to assume the space $\times$ time factorization: $M \sim \Sigma \times {\mathbb R}$, such that $X^i, i=1,...d-1\; \in\;\Sigma$, and $X^0\,\in\, {\mathbb R}$. Then we have a spacetime foliation $\Sigma(X^0)$, and by fixing one of the string coordinates $\sigma^0 = X^0 (\equiv t) $, we can perform a standard $d+1$ decomposition of the intrinsic variables $(q_{ab}(\sigma^c),\phi(\sigma^c))$, and then define the canonical Hamiltonian $H$.

The foam $W$ intersects the spacial sheet $\Sigma(t)$ in a set of $N(t)$ smooth (open or closed) oriented curves $\gamma_{1 \leq i \leq N}:[0,\pi] \to \Sigma\,$, $\Gamma(t)\equiv X(W) \cap\Sigma(t)=\cup^{N(t)}_i \,\gamma_i\,$, which properly resembles a classical system of strings.

%\vspace{0.4cm}

\subsection{The Canonical Ensemble}

%\vspace{0.4cm}

Let us suppose that we only have some information on the internal energy of the string foam, namely, the averaged energy $\langle H\rangle$ is a precise number: $u$. In this ensemble, it is implicitly assumed that the system of strings may exchange energy with the exterior through of the boundary $\partial X(W) \equiv \partial M \cap X(W) $ (via contact interactions), such that the mean value
\be\label{constr-m}
u(\Sigma)=\langle H \rangle= Tr \r H
\ee
remains fix among macroscopic observations. In other words, this quantity shall not depend on the macroscopic instruments, associated to the observer affine parameter $X^0$.
This is a macroscopic observable, which  may be identified as the total mass-energy enclosed in the box $\Sigma$, and
the Helmholtz free energy is defined by
\be\label{Fus}
F (\Sigma, \b)= u - \b^{-1} s\,.
\ee
To follow the standard procedure, we must maximize the entropy functional with respect to $\r$, keeping the constraint (\ref{constr-m}) and the normalizing condition,
\be\label{tr}
Tr \r =1\,\,,
\ee where the trace is taken over ${\cal H}_{(\z,W)}$. So one have the equation (we set the Boltzmann constant $k_B=1$) :
\be\label{Smin}
\delta \,\, Tr \left( \gamma \r + \beta H \r  - \r \ln \r \right) =0 \, ,
\ee
then
\be\label{Frho}
 \,\, \left( -\gamma + \beta H  +  \ln \r + 1 \right)  =0 \,,
\ee
where $\gamma, \b$ are lagrange multipliers, that for an homogeneous/uniform system do not depend on the space point \footnote{This issue will be more carefully discussed in Sec. 5.}.
The density operator then results
\be\label{rho}
  \r = e^{\left( \gamma -1\right)} \,\,e^{ -\beta H} = Z^{-1} \,\,e^{ -\beta H} \,.
\ee
The partition function is defined as
\be\label{Z}
Z= e^{\left( 1- \gamma\right)} = Tr \, e^{-\beta H}\,\,,
\ee
which by virtue of eq. (\ref{Frho}), may be expressed in terms of the free energy
\be\label{defF}
Z(\Sigma, \b) \,= \, e^{-\beta F}\,\,.
\ee
By using a well-known trick, this may be written as a path integral of the action on the analytical continuation to the foam of two-dimensional Euclidean worldsheets $\widetilde{W}$:
\be\label{Zcan1} Z(\Sigma, \b)= \sum_{\widetilde{W}}\,
\int_{{\cal F}_{int}/G}  {\cal D} \tilde{q} {\cal D} \phi \;\;  e^{-{S}[\tilde{q}, \phi]} \; , \ee
 where the sum is over all the Riemannian compact geometries with Euclidean metric $\tilde{q}$. The topology of $\widetilde{W}$ may be expressed as $ \Gamma \times S^1_\beta$ (where $S^1_\beta$ denotes the circle of radius $\b/2\pi$).
This expression is more convenient for the partition function because it sums over the classical microscopic geometries
 and captures the fundamental symmetries of the theory \cite{haw-gib, EQGbook}.
${\cal D} \tilde{q} {\cal D} \phi$ schematically denotes
the appropriate measure on the space of orbits ${\cal F}_{int}/G$,
where ${\cal F}_{int}$ is the space of all the intrinsic fields $(\tilde{q}, \phi) $ on $\widetilde{W}$ (the embedding field $X$ was integrated out and absorbed in a normalizing factor), and the gauge group $G$ consists of
worldsheet diffeomorphisms and Weyl transformations of $\widetilde{W}$, and the internal gauge group associated to $\phi$.
%\textbf{footnote: In this ensemble, the embedding fields $X^i(\sigma)$ is assumed fixed (up to $Diff(M)$) }
Notice furthermore that the string picture is non-perturbative in this framework, and the thermal partition function above is exact.

The topologies of $\widetilde{W}$ are totally characterized by the genus $g$ and the number of boundaries or holes $h$, so the sum above may be expanded in integers powers of the parameter $e^{\lambda}$. The geometry action is $S_G[\tilde{q}] = \lambda \chi$, where $ \chi= 2-2g-h $ is the Euler number,
then the canonical partition function (\ref{Zcan1}) can be expressed as
\be Z(\Sigma, \b)= \sum_{g,h}\,\, e^{-\lambda\, \chi }\,
\int_{{\cal F}_{int}/G}  {\cal D} \tilde{q} {\cal D} \phi \;\;   e^{-{S}_m[\tilde{q}, \phi]} \,. \ee
%where ${\cal V}\equiv Diff (\widetilde{W}_\b) \times Weyl(\tilde{q}) \times Gauge(\phi)$
Let us stress that in the present approach, $Z$, must be strictly interpreted as a traditional partition function of statistical mechanics, rather than as the Wick-rotated Feynman path integral formulation of the quantum theory; although the definition $\langle {\cal O}\rangle \equiv Tr_{(\cal H)} \,\r\,{\cal O}$  involves both, quantum and statistical average.
On the other hand however, the quantization of theory (\ref{action}) may be defined in the Feynman's picture, and then the
transition amplitudes and $n$-point correlation functions may be computed by properly inserting operator products in the above path integral. Notice furthermore that in principle this string model is non-perturbative, and the thermal partition function above is exact.

We wish emphasize finally that the partition function is the most important object in the present construction since it
encodes all the thermodynamical (and quantum) information of the system, and in particular the spacetime physics must be extracted from this. This fact resembles the Euclidean quantum gravity viewpoint, whose expectation is that the sum over Euclidean paths should play a fundamental role in the description of the spacetime geometry \cite{EQGbook}.

\vspace{0.7cm}

As an illustrative example, we may easily evaluate the thermodynamical entropy associated to the space geometry $\Sigma$ in the canonical ensemble. In the simplest microscopic theory, where the contribution of local degrees of freedom is neglected ($S_m \approx 0$), the internal energy and the entropy are:
\be
u(\Sigma) =\frac{\partial\, ln\, Z}{\partial \beta}= 0 \;\; , \;\;\;\; s(\Sigma)= ln \, Z \approx ln \, \left( \sum_{g,h}\,\, e^{-\lambda\, \chi }\,\right).
\ee
If we only consider closed strings, $h$ is given by the number of intersections of $\Gamma$ with the boundary of the space $\partial \Sigma$. So if this number is fix (as a boundary condition), we obtain a linear law,
\be
s(\Sigma) \approx \lambda h\, +\, ln \, \left( \sum_{g}\,\, e^{2\lambda\, (g-1)}\,\right),
\ee
relating the entropy and the number of boundaries of the string lattice.

%\vspace{0.4cm}

\subsection{Generalized Ensembles}

%\vspace{0.4cm}

The canonical ensemble may be generalized to other ones by fixing certain macroscopic quantities. Let us consider first a collection (labeled by the index $I$) of local operators  of the theory
\be\label{localope}
{\cal C}^I (\z(\sigma^a)) \, ,
\ee
which are gauge invariant pointwise functions of the microscopic fields $\z(\sigma^a)=(X^\mu(\sigma^c), q_{ab}(\sigma^c),\phi(\sigma^c))$ (and its derivatives). Then we can define extensive quantities (observables) by integrating this on the string lattice\footnote{In this expression $\Gamma$ is used to denote the points of the string foam (a set of curves) whose embedding lies on $\Sigma(t)$. }.
\be\label{observable}
{\cal C}^I(\Sigma)= \int_{\Gamma} \,\,\sqrt{q_{11}}\,\,{\cal C}^I(\z(\sigma^a))\,\,\, d\sigma   .
\ee
Notice is that in principle the label index $I$ may run over objects that transform as tensors of $M$, so it may enclose indices $\mu, \nu$. The averaged operators
\be\label{constraints}
C^I(\Sigma)\equiv\langle{\cal C}^I (\Sigma)\rangle= Tr \,\r \,\, \,{\cal C}^I(\Sigma) ,
\ee
are accessible in macroscopic measurements. If one assumes that these macroscopic quantities are constant, then the free energy is defined as
\be\label{omega-fund}
\O(\Sigma, \b, \mu_I) = u - \b^{-1} s - \sum_I \;\mu_I \, C_{I}[\Sigma]\ee
where $\mu_I$ are the potentials (lagrange multipliers, in statistical language), associated with the quantities macroscopically conserved (\ref{constraints}). By maximizing the functional of entropy with these constraints (\ref{constraints}), the probability density operator and free energy become:
\be \r =  \,{\cal Z}^{-1}\; \exp^ {-\b \left( H  - \sum_I \;
\mu_I \, C_{I} \right) }\; , \label{granrho}\ee
\be\label{omega} \O= - \b^{-1}\,\ln {\cal Z} \, ,\ee
where the partition function expresses:
\be {\cal Z}= \; Tr \; \exp^ {-\b \left( H - \sum_I \;
\mu_I \, C_{I} \right) }\; . \label{granZ}\ee
This may be expressed in terms of sum over Riemman surfaces as:
\be {\cal Z}(\Sigma, \b, \mu_I)=
\sum_{\widetilde{W}}
\int_{{\cal F}/G} {\cal D} \tilde{X} {\cal D} \tilde{q} {\cal D} \phi \;\;  e^{-\left( {S}[\tilde{q}, \phi] - \sum_I \;
\mu_I \, \tilde{C}^I \right)} , \ee
where the path integral involves the global quantity
\be\label{observableglobal}
\tilde{C}^I \equiv \int_{\widetilde{W}} \,\,\sqrt{\tilde{q}}\,\,\tilde{{\cal C}}^I \,\,\, d\tilde{\sigma}^2   .
\ee
where $\tilde{{\cal C}}^I \, \equiv \, {\cal C}^I(\tilde{\z})\,$ are the local operators (\ref{localope}) defined on the Euclidean section of the analytical continuation of the variables: $\tilde{\z}=(\tilde{X}(\tilde{\sigma}^c), \tilde{q}_{ab}(\tilde{\sigma}^c),\phi(\tilde{\sigma}^c))$ as usual. Therefore the macroscopic measures (\ref{constraints}) may be computed through insertions of these operators in the path integral as
\be\label{constraints-path}
C^I(\Sigma)= -\frac{\partial {\Omega}}{\partial \mu_I}= -\frac{1}{\b {\cal Z}}\sum_{\widetilde{W}}
\int {\cal D} \tilde{X} {\cal D} \tilde{q} {\cal D} \phi \;\; \tilde{ C }^I\;\; e^{-\left( {S}[\tilde{q}, \phi] - \sum_I \;
\mu_I \, \tilde{ C}^I \right)} .
\ee
These are the equations of state that relate the quantities $C^I(\Sigma, \beta, \mu_I)$ to the potentials $\mu_I$.

\subsection{The Geometric Ensemble}

The choice of the ensemble is crucial to extract desiderated macroscopic information, which obviously depends on the choice of what macroscopic quantities are fixed. In order to recover information on the extrinsic geometry of the string-foam and the spacetime geometry itself,
let us consider the local operator:
\be\label{thetalocal}
\Theta^{\mu\nu}=\, \frac{1}{4\pi\alpha} \, q^{a b} \partial_b X^\nu
\partial_a X^\mu\,,
\ee
defined in terms of the microscopic variables, where the embedding sector enters explicitly.
Integrating this on the strings lattice, we obtain the extensive observable:
\be\label{theta}
\Theta^{\mu\nu}[\Sigma]=\, \frac{1}{4\pi\alpha} \,\int_{\Gamma} \,\,\sqrt{q_{11}}\,\,\, d\sigma \,\,\, q^{a b} \partial_b X^\nu
\partial_a X^\mu \,.
\ee
This object is the natural generalization to strings of the microscopic energy-momentum tensor of a perfect fluid (see next Section). The constant $\a$ is introduced to give correct unities of energy density.

The crucial point of our construction is to define the \emph{geometric ensemble} by assuming the following macroscopic quantity
\be\label{thetaTr}
\theta^{\mu\nu}[\Sigma]\equiv \langle\Theta^{\mu\nu}[\Sigma]\rangle= Tr \,\,\r\,\, \Theta^{\mu\nu}
\ee
to be conserved. So we are going to consider a particular ensemble where the \emph{averaged} spacial distribution of the energy-momentum of the system of strings is fix.

%\textbf{[fr1]- If the system is assumed to be "$\Theta$-isolated", i.e, such that it does not exchanges this quantity with another external system (reservoir), then normal thermodynamic arguments led to the equilibrium value $\theta^{\mu\nu} \approx 0$. This is the case we are going to consider here for simplicity.}
%Dbranas! son la otra posibilidad
%This essentially means that the macroscopic quantity number (energy density of the embedding fields) $\epsilon_U \equiv UU\Theta$
%is fix and independent of $U$
%\textbf{Thermodynamic stability arguments require that $g$ be positive definite, which means that the spacetime metric is Euclidean.}

The case we consider first is an isotropic and uniform mean distribution: $\theta \approx constant$. As shown before, one shall to introduce a new set of Lagrange multipliers (potentials), corresponding to this constraint, namely: \textbf{$\mu$} $\equiv\, g_{\mu\nu}$. Then, by virtue of (\ref{granZ}) the grand partition function writes
\be {\cal Z}(\Sigma, \b, g_{\mu\nu}) = \; Tr \; \exp^{-\b \left( H - \; \sum_{\mu \nu}
g_{\mu\nu} \, \Theta^{\mu\nu}\right) }\; \label{granZ-geom}.\ee
Invariance of this expression under spacetime diffeomorphisms, implies that $g_{\mu\nu}$ must transform as a symmetric rank-two tensor of $M$. Therefore, this may be properly identified with a \emph{metric} on $M$
\footnote{Thermodynamical stability arguments, and positive-definiteness of $\Omega$ require that $\tilde{\theta}^{\mu\nu}$ be has
 the same signature that $g_{\mu\nu}$, which is positive definite, and so the thermodynamical spacetime metric is Riemannian.
  However, when one looks for the microscopic quantum field theory, $q_{ab}$ is Lorentzian, and then $g_{\mu\nu}$ should
   be consistently continued to the Lorentzian section.}.
%and interpreted as effective field.}.
\be\label{granZ-gpath} {\cal Z}(\Sigma, \b, g_{\mu\nu})=
\sum_{\widetilde{W}}
\int {\cal D} \tilde{X}  {\cal D} \tilde{q} {\cal D} \phi \;\;  e^{-\left( {S}[\tilde{q}, \phi] - \,
g_{\mu\nu} \, \tilde{ \theta}^{\mu\nu} \right)} , \ee
where
\be\label{thetaglobal}
\tilde{\theta}^{\mu\nu} \equiv \, \frac{1}{4\pi\alpha} \,\int_{\widetilde{W}} \,\,\sqrt{\tilde{q}}\,\,\tilde{\Theta}^{\mu\nu}(\tilde{\sigma}^a) \,\,\,  d\tilde{\sigma}^2  = \, \frac{1}{4\pi\alpha} \,\int_{\widetilde{W}} d\tilde{\sigma}^2 \,\, \tilde{q}^{\frac{1}{2}}\,\,\tilde{q}^{a b} \partial_b X^\nu
\partial_a X^\mu \; .
\ee
Assuming now macroscopical uniformity of the system: $g_{\mu\nu}\approx constant$, we can plug this into the integral of the exponential (\ref{granZ-gpath}), and obtain the thermodynamical partition function
\be\label{Poly-flat} {\cal Z}(\Sigma, \b, g_{\mu\nu})=
\sum_{\widetilde{W}}
\int {\cal D} \tilde{q} {\cal D} \phi \;\;  e^{-\left( S[\tilde{q}, \phi] -  \, \, \frac{1}{4\pi\alpha} \,\int_{\widetilde{W}} d\tilde{\sigma}^2 \,\,g_{\mu\nu}\, \tilde{q}^{\frac{1}{2}}\,\tilde{q}^{a b} \partial_b X^\nu
\partial_a X^\mu \right)} . \ee

By taking the simplest microscopic theory where intrinsic fields ${\cal L}_m = 0$, apart from those that describe the geometry are absent. This remarkably coincides with the Polyakov partition function of interaction-full string theory \cite{jpbook}. The background metric play the role of a thermodynamical potential, which by first principles, implies that it is a c-number and should not be quantized as believed in standard string theory.

Although this approach is mathematically very simple, it is not obvious that a constraint on the mean extrinsic geometry of the string foam $W$, leads to introduce an ambient spacetime metric as a Lagrange multiplier.
Let us notice that in the present framework, the
%only difference between the topological theory (Eq. (\ref{action})), and the physical string one (Eq. (\ref{Poly-flat})), is encoded in the ensemble choice.
only difference between the topological theory (Eq. (\ref{action})), and the physical string one (Eq. (\ref{Poly-flat})), is encoded in the ensemble used
\footnote{While in the canonical ensemble we do not local observables, in the geometric one, there are (macroscopical) quantities localized in the spacetime manifold, involving the embedding fields (Eq. (\ref{thetaTr})).}.
We speculate that many apparently different string theories might be related through a kind of Legendre transformation relating different ensembles.

%\vspace{0.3cm}

\subsection{Thermodynamic equilibrium: Einstein equation of state}

Let us consider now that the potential $g_{\mu\nu}$ may depend on the
 spacetime point; it shall be discussed in the next Section.

In the thermodynamic limit, one must integrate out the microscopic degrees of freedom such as prescribes (\ref{Poly-flat}), so the thermodynamical partition function only depends on the thermodynamic variables:
\be\label{effective} {\cal Z}(\Sigma, \b, g_{\mu\nu})= \;\;  e^{- I_{}[g] } , \ee
which to first order in the parameter $\a$, is known to be \footnote{Here for simplicity, we shall assume critical dimension $d=26$ and ignore other spacetime fields.}:
\be
I_{}[g] = I_{EH}[g] + o(\a) \,\,\,\,, \,\,\,\,\,\,I_{EH}[g]= \frac{1}{16\pi} \int_{\Sigma
\times S^1_\b} dx^{d} \; \sqrt{g} \; R(g)  .
\ee
By virtue of (\ref{omega}), this quantity is related to the thermodynamical free energy for this ensemble
\be\label{fundamental}
\Omega(\Sigma, \b, g_{\mu\nu}) \equiv I / \b = \frac{1}{\b}\left(\frac{1}{16\pi}\int_{\Sigma
\times S^1_\b} dx^{d} \; \sqrt{g} \; R(g)\right) + o(\a)\,.
\ee
The normalizing constant $1/16\pi$ may be changed by a redefinition of the inverse temperature parameter $\beta$.
 Notice that, in this approach, the role played by $\beta$ is similar in some
 aspects to the dilaton in conventional string theory, which suggests
 an interesting interpretation for this background field; it shall be explored
 elsewhere.

 On the other hand, consistently with this argument, the analytical
 continuation to Lorentzian time of the partition function (\ref{Poly-flat})
 (and the correlation functions) \emph{defines} the path integral quantization
 of the microscopic theory, which of course coincides with conventional string
 theory. Then, the known analysis of the beta-functions even requires that the
 Einstein equations be satisfied to leading order in $\a$.

In the thermodynamical language highlighted in this approach, (\ref{fundamental}) is nothing but the \emph{fundamental relation} in the geometric ensemble; therefore, the equations of state are given by equating the quantities (\ref{thetaglobal}) to the variation of $\Omega$ with respect to the thermodynamical potentials. To leading order in $\a$ and constant $\b$, we obtain:
\be\label{einstein-theta}
\theta^{\mu\nu} = R^{\mu\nu}[g]- \frac12 \, R[g]\,g^{\mu\nu}  \,\,.\ee
This is the Einstein equation, recovered here as an \emph{equation of state} in agreement with the Jacobson's paradigm.

In absence of other background fields (interpreted here as thermodynamic variables), $\theta^{\mu\nu}$ is only a function of $\beta$, and $g_{\mu \nu}$; in the maximally uniform case, this just can be $\theta^{\mu\nu} \sim \Lambda(\beta, det (g) ) \, g^{\mu\nu}$, where $\Lambda$ is a global quantity, which may be suggestively expressed in terms of the microscopical variables as: \be
\Lambda \sim \, \langle \, \frac{1}{4\pi\alpha} \,\int_{\widetilde{W}} d\tilde{\sigma}^2 \,\,g_{\mu\nu}\, \tilde{q}^{\frac{1}{2}}\,\tilde{q}^{a b} \partial_b X^\nu
\partial_a X^\mu \, \rangle .\ee
 For simplicity and consistency, in what follows, we constrain this quantity
 to be zero, and fix a vanishing macroscopical constraint (\ref{thetaTr}) in general.
 Then, the equation of state becomes:

%For simplicity and consistency, in what follows we constraint this quantity to be zero, and fix a vanishing macroscopical constraint  in general.
%\textbf{In order to avoid infinities..... $\Lambda=0$......}
%\be\label{thetaTr=0}
%\theta^{\mu\nu}[\Sigma] =  Tr \,\,\r\,\, \Theta^{\mu\nu} = 0\,
%\ee
Then the equation of state becomes:
\be\label{einstein}
 R^{\mu\nu}[g] = 0 \, ,\ee

A solution to this equation is uniquely determined by Dirchlett boundary conditions, as happens for an intensive potential in standard thermodynamics (e.g. the temperature distribution in a material medium).
For metrics which satisfy (\ref{einstein}) the geometric free energy $\O$ is an extremum under arbitrary variations
 of $g_{\mu\nu}$, whose value must be fix on the boundary; so for consistency, this must be supplemented with a boundary term:
%\be
%\label{Free-EH} I[g_{\mu\nu}] = \int_{\Sigma
%\times S^1_\b} d^4 x \; \sqrt{g} \; R(g) +
%\int_{\partial ( \Sigma \times S^1_\b )} K[g]\; + \;o(\a)\;,\ee
%Then the geometric free energy becomes
\be
\label{Free-EH} \Omega(\Sigma, \b, g_{\mu\nu}) = \frac{1}{\b}\left(\frac{1}{16\pi}\int_{\Sigma
\times S^1_\b} d^4 x \;  \sqrt{g}\; R(g)\;+\; \frac{1}{8\pi}\int_{\partial ( \Sigma \times S^1_\b )} \sqrt{h} \,  (K + C) \right) + o(\a)\,.
\ee
where $K$ is the trace of the second fundamental form of the boundary in the metric g and $C$
 depends only on the induced metric $h$ on $\partial ( \Sigma \times S^1_\b )$. This term may be absorbed in a proper redefinition of the zero free energy, so
 we can substitute $K+C$ by $\Delta K$, the difference in the trace of the second fundamental form of the boundary in the metric $g$ and another flat $g^0$.

We can put the solution in terms of the boundary conditions and plug this back into the expression (\ref{Free-EH}). By using the equation of state (\ref{einstein}), we eliminate the potential $g$ and so the free energy reduces to the Helmholtz's potential:
\be
\label{FG} F(\Sigma, \b) =  \frac{1}{8\pi}\int_{\partial ( \Sigma \times S^1_\b )} \sqrt{h} \, \Delta K\;= \,\Delta\,\frac{\partial}{\partial \mathbf{n}} \int_{\partial ( \Sigma \times S^1_\b )} \sqrt{h} .\ee
This expression is very useful to evaluate the entropy and other thermodynamical quantities in the present framework.
 In fact using the formulas above, standard thermodynamical relations and an explicit solution to the equation of state (\ref{einstein}), the computation
  procedure is rather similar to that developed in the context of Euclidean quantum gravity (Ref. \cite{haw-gib, EQGbook}),
   although here it is based on very different arguments \footnote{In particular, the metric is treated now as a thermodynamical potential
    and the derivation involves a macroscopic limit rather than a semiclassical/saddle point approximation.}.
 %\textbf{although reproduced here in a radically different construction.}

 Let us briefly illustrate this in the example of a Schwarzschild solution describing a black hole.
 Start with a spacetime manifold $M=M_4 \times K_{d-1}$ where $K_{d-1}$ is compact (let us ignore it here for simplicity), and as assumed above $M_4 \sim \Sigma_3 \times S^1_\b$. Let us assume also spherically symmetric boundary conditions such that the potential $g$ is constant on the boundary of $\Sigma_3$ whose topology is assumed to be $\sim S^2$. Then the solution of (\ref{einstein}) is the Euclidean section of the analytically continued Schwarzschild solution,
 \be\label{solution}
 ds^2 = (1-2M/r) d\tau ^2 + (1-2M/r)^{-1} dr^2 +  r^2 d\Omega^2 \,\,\,(+\, \,ds_K^2)\,\, ,   \,\,r \geq 2M\,\,
 \ee
 where $\tau$ is the coordinate of $S^1_\b$. The coordinate singularity in $r=2M$ may be removed by defining a new radial coordinate $x\equiv 4M (1-2M/r)^{1/2}$.

 In order to have a free energy $\Omega$ finite, the parameter $M$ must be equaled to $\b / 8\pi$. Therefore the
 the formula (\ref{FG}) (neglecting $o(\a)$ corrections) may be straightforwardly evaluated \cite{haw-gib}:
 \be
 F= \frac{\beta}{16\pi} (= M/2) + o (r_0^{-1}),
\ee
where $r_0$ is the radial coordinate of the boundary sphere. From the standard thermodynamical expression:
\be
u= -\frac{\partial (\beta F)}{\partial \beta} =  \frac{\beta}{8\pi} \,\,,
\ee
and using that $F = u- s/\b$, we obtain the entropy:
\be
s= \frac{\beta^2}{16\pi} \,.
\ee
The area of the horizon surface $r\equiv 2M$ is $A=16\pi M^2$ by virtue of  (\ref{solution}). Then we recover the Beckenstein-Hawking formula $s= A/4\,$.

Notice that in conventional string theory, known black hole entropy calculations address to compute this in the \emph{microcanonical} ensemble
 by counting the number of microscopic $D$-brane configurations (defined on flat background) \cite{vafa-strom}. In this model we emphasize that this is not necessary, and in contrast, a very different canonical ensemble (geometric) should be used to compute these quantities, because of the very thermodynamical meaning of the background geometry.

%Two comments are in order: i)
%Three remarks are in order: i)
%Then we achieved two very important results of the present approach:

Let us summarize some remarks about the results we have achieved so far:

\begin{itemize}

 \item Einstein equation was recovered as equilibrium condition in the limit $\a\to 0$, that is, as equation of state. This agrees with the perspective highlighted by Jacobson \cite{jacobson}.

%Helmholtz free energy is recovered by eliminating the variable g(x) from its equilibrium value

\item Corrections in the parameter $\a$ must be interpreted in this picture as describing local deviations from the thermodynamic equilibrium/uniformity of the strings fluid. In fact, a non-negligible $\a$-scale mean that the microscopic details of the geometry (strings) manifests.
 So we can claim that this parameter \emph{controls} the macroscopic limit: macroscopic length scales correspond to very small string
  length compared with the measurements of space-time intervals.
    %\textbf{Equivalently, very large number of strings filling $\Sigma$, forces $\alpha$ to be small compared with the total $Vol(\Sigma)$.}

 \item Gauss-Bonnet terms \cite{gross-witten} appear in higher order  $\a$-corrections of (\ref{Free-EH}). Since it describes deviations from the local thermodynamic equilibrium, we can expect that an hydrodynamic description of the spacetime corresponds to higher order corrections of Einstein gravity. This is in complete agreement with generalizations of the Jacobson result \cite{jacobson-noneq}

     \item Black holes entropy may be recovered in this approach. Even though the derivation resembles the Euclidean quantum gravity procedure, it must be clarified that in this approach the thermodynamical state is naturally described by an euclidean solution of the Einstein equations, and we do not need to assume a saddle-point/semiclassical approximation (in place of that, a macroscopic limit), nor any artificial interpretation of the imaginary time.

\end{itemize}

\section{Local constraints and the string $\sigma$-model}

In the previous section we have assumed that the string scales are small compared with the scale of lengths used by the observer.

In order to generalize the above model to general curved background and recover the full string theory sigma model, we have to slightly modify the macroscopic constraint and take into account the dependence with the spacetime point. This scales are assumed to be sufficiently short to see local variations of the spacetime (macroscopic) observables but large enough to be described as a thermodynamical system.

As an analog example, consider a relativistic perfect fluid whose microscopical components are structureless point like particles that interact only in spatially localized collisions in a compact region of the space ${\cal V} \subset \Sigma$. For any point $x^i \in {\cal V}$, the energy-momentum tensor reads
\be\label{Tpoint}
T^{\mu\nu}(x^i)= \sum_{n=1}^N \,\frac{m}{2} \,\partial_t X_n^\mu \partial_t\, X_n^\nu \,\,\delta^{d-1}(x^i-X^i_n)\,\,\,
\ee  where $m$ is a constant mass parameter and $X^0=t$.
In this case the microscopical variables are the positions of the individual particles $X_n(t)$. The particle number density is
\be\label{npoint}
n(x^i)= \sum_{n=1}^N \,\,\delta^{d-1}(x^i-X^i_n)
\ee
Taking the limit $N\to\infty$ in a fix volume $V \equiv Vol({\cal V})$, this becomes a continuum with an uniform distribution of particles. This will be the case if the mean distance between particles (or the mean free path between collisions) is small compared
 with the scale of lengths used by the observer.
 So the sum above becomes
\be\label{npoint-limit}
n(x^i) \sim \int_{{\cal V}} dX^i \,\,\delta^{d-1}(x^i-X^i)\,\, .
\ee
By taking the same limit in the expression (\ref{Tpoint}), we recover the energy-momentum tensor field of a (continuum) relativistic fluid, which reads:
%($N\to \infty , V =constant$) in the expression (\ref{Tpoint}),
%in this (hydrodynamic) approximation, the energy momentum tensor reads
\be\label{Tpoint-limit}
T^{\mu\nu}(x^i)= \int_{{\cal V}} dX^i \,\,\frac{m}{2} \,\partial_t X^\mu \partial_t\, X^\nu \,\,\delta^{d-1}(x^i-X^i)\,.
\ee
%\bea\nonumber\\&=&\,\frac{m}{2} \,(\, \partial_t x^i \partial_t\, x^j \delta_i^\mu \delta_j^\nu + \partial_t\, x^j \delta_0^\mu \delta_j^\nu + \partial_t x^i \,\delta_i^\mu \delta_0^\nu + \delta_0^\mu \delta_0^\nu )\,.\eea
%Our string foam is quite similar to this system of colliding particles, which gives us a good insight into our approach.
Inspired in equation (\ref{Tpoint}), let us define the string operator:
%\be\label{constr-sigma}
%\Theta^{\mu\nu}(x^\a)=\,\int_{\Gamma} \,\,\sqrt{q_{11}}\,\,\, d\sigma \,\,\, q^{a b} \partial_b X^\nu\,\partial_a X^\mu\,\,\delta(x^i - X^i(\sigma^a))\,\delta(x^0 - t)\,
%\ee
\be\label{constr-sigma}
\Theta^{\mu\nu}(x^\a)=\, \, \frac{1}{4\pi\alpha} \,\int_{\Gamma} \,\,\sqrt{q_{11}}\,\,\, d\sigma \,\,\, q^{a b} \partial_b X^\nu\,\partial_a X^\mu\,\,\delta(x^\a - X^\a(\sigma^a))\,\,,
\ee
for each spacetime point $x^\a $. In the string system we do not have an obvious notion of $N$, but in the continuum/uniform limit may be thought as the string lattice $\Gamma$ (almost) filling the whole compact space $\Sigma$ (schematically, $\Gamma \to \Sigma$), then the point $x^\a$ lies on $\Gamma$ and then this operator coincides with (\ref{theta}) defined before.

The generalization of the constraint (\ref{thetaTr}) is
\be
\theta^{\mu\nu} (x^\a) = Tr \,\,\r\,\, \Theta^{\mu\nu} (x^\a) = constant (\equiv 0)
\ee
Following the same procedure of Section \textbf{3.2}, we must introduce Lagrange multipliers (potentials) corresponding to this constraint, that depend on the spacetime point: $g_{\mu\nu}(x^\a)$. Then, by virtue of (\ref{omega-fund}) we have:
\be
\O(\Sigma, \b, g_{\mu\nu}(x)) = u - \b^{-1} s - \; \int_\Sigma dx\,\,
g_{\mu\nu}(x) \, \Theta^{\mu\nu}(x) \,\,;\ee
and the grand partition function expresses:
\be {\cal Z}(\Sigma, \b, g_{\mu\nu}(x)) = \; Tr \; \exp^{-\b \left( H - \; \int_\Sigma dx\,\,
g_{\mu\nu}(x) \, \Theta^{\mu\nu}(x)\right) }\;  \label{granZ-geom-x}.\ee
%\textbf{duda: va sigma o M?}
So now, by generalizing the uniform tensor (\ref{thetaglobal}) to
\be\label{thetaglobal-x}
\tilde{\theta}^{\mu\nu}(x^\a) \equiv \,\int_{\widetilde{W}} \,\,\sqrt{q}\,\,\tilde{\Theta}^{\mu\nu}(\sigma^a) \,\,\,  d\sigma^2  = \, \frac{1}{4\pi\alpha} \,\int_{\widetilde{W}} d\sigma^2 \,\, \tilde{q}^{\frac{1}{2}}\,\,\tilde{q}^{a b} \partial_b X^\nu
\partial_a X^\mu \,\,\delta(x^\a - X^\a(\sigma^a))\,\,,
\ee
the grand partition function may be written as a sum over two-dimensional geometries in terms of an Euclidean action, which looks like a non-linear sigma model:
\be\label{Z-sigma} {\cal Z}(\Sigma, \b, g_{\mu\nu})=
\sum_{\widetilde{W}}
\int {\cal D} \tilde{q} {\cal D} \phi {\cal D} X \;\;  e^{-\left( {S}[\tilde{q}, \phi] -  \,\, \frac{1}{4\pi\alpha} \, \int_{\widetilde{W}} d\sigma^2\,g_{\mu\nu}(X^\a(\sigma^a))\, \,\sqrt{\tilde{q}}\,\,\,\tilde{q}^{a b} \partial_b X^\mu
\partial_a X^\nu \right)}\,\, . \ee
Once again, by taking the intrinsic action $S=\lambda S_G$, this is the string theory Polyakov partition function on general spacetime metrics, described here as point-dependent thermodynamical potentials.

%\textbf{(Notice that in eq. (\ref{constr-sigma}) we relax the spacial localization condition, otherwise if we impose it, we would recover only static spacetime metrics)}

So in general, in order to describe non-uniform thermodynamical constraints in our model, we must to generalize Eqs. (\ref{observable}) by considering operators \be\label{constraints-x}
C^I(x) = \int_{\Gamma} \,\,\sqrt{\tilde{q}}\, d\sigma^2 \,\,\,  {\cal C}^I(\z(\sigma^a))\,\,\delta(x^\a - X^\a(\sigma^a))\,,
\ee
whose expectation value is a conserved quantity. Then, the most general grand canonical partition function writes:
\be {\cal Z}(\Sigma, \b, \mu_I(x) )=
\sum_{\widetilde{W}}
\int {\cal D} \tilde{q} {\cal D} \phi {\cal D} X\;\;  e^{-\left( {S}[\tilde{q}, \phi] -  \, \int_{\widetilde{W}} d\sigma^2\,\sqrt{\tilde{q}}\,\,\,\mu_{I}(X^\a(\sigma^a)) \tilde{{\cal C}}^I(\z(\sigma^a)) \right)} , \ee
where $\tilde{{\cal C}}^I$ is the local operator (\ref{localope}) defined on the Euclidean section of the analytical continuation $\widetilde{W}$ as before.
The simplest meaningful operator of this type one can define is:
\be\label{n-string}
\mathcal{N} (x^\a) = \,\int_{\Gamma} \,\,\sqrt{q_{11}}\,\,\, d\sigma \,\,\,\delta(x^\a - X^\a(\sigma^a))\,,
\ee
which simply integrates the intrinsic volumes of the strings distribution, it is similar to (\ref{npoint}) and may be interpreted as a string density. In a closed system this should be a fix (very large) number, while in a more natural open system described by a grand canonical ensemble, this would contribute to the path integral (\ref{Z-sigma}) with a term in the exponential:
\be
\int_{\widetilde{W}} \, \,\sqrt{\tilde{q}}\,\,\,d\sigma^2 \,\,\mu(X^\a(\sigma^a))\,\;,
\ee
where spacetime field $\mu$ is the chemical potential.

\section{Concluding remarks}

We have introduced a microscopic model for the spacetime, which is viewed as a macroscopic system governed by the statistical mechanics and thermodynamics laws. The picture is a large number of two-dimensional base manifolds: an evolving string lattice.
We have shown that the metric and other spacetime fields shall not be quantized, but treated as ordinary thermodynamical quantities (temperature, pressure, chemical potential, an so on). The geometric ensemble was introduced and the Einstein equation was recovered as an equation of state, in a remarkable agreement with the Jacobson paradigm.

The model matches string theory, but implies some changes in its formulation and in the way that it should be understood; in particular:
 i) the microscopic (string) theory is fundamental and non-perturbative; ii) its formulation is background-independent in the usual sense;
 iii) the constant $\a^{1/2}\,(>l_P)$ is associated to the microscopical (although non-Planckian)
  length scales and controls the macroscopic (long wavelength) limit; corrections in $\a$ should be related to transport phenomena and hydrodynamic description;
%the meaning some aspects of string thermodynamics (e.g. the Hagedorn bound) should be revisited
 iv) the relation between different string theories, and dualities, should be reassessed under the perspective of statistical ensembles; and
 v) it constitutes a strong input on the description of the spacetime physics.

 This last issue is particularly meaningful since the quantum gravity problem reduces to exactly quantizing a two-dimensional theory, which may be completely solved in a context consistent with string theory. According to this perspective, many topological properties of the spacetime shall be restricted, and furthermore, changes of topology, compactifications, and even gravitational collapse, should be described as thermodynamical processes. These subjects shall be studied in forthcoming works.

\section{Acknowledgements}

 The author is grateful to the Theory Unit at CERN
 for providing a comfortable and stimulating work environment.
   L. Alvarez-Gaume is specially acknowledged for interesting comments and
 observations. Thanks are due to J. A. Helayel-Neto for reading the manuscript and
 A.L. Nogueira for stimulating comments.
 This work was partially supported by: CONICET PIP 2010-0396 and ANPCyT PICT 2007-0849.
%The author is grateful to colleagues of the Theory Unit at CERN
% during the preparation of the manuscript.
%> Thanks are due to  J. A. Helayel-Neto for reading the manuscript and
%> stimulating observations.
%> XXXX is specially acknowledged for specially useful comments and
%> observations.
%> CONICET is acknowledged for financial support.
%> This work was partially supported by grants: ANPCyT PICT 2007-0849 and
%> CONICET PIP 2010-0396.

\end{document}